\documentclass[conference,a4paper]{IEEEtran}
\usepackage{cite}
\usepackage{amsmath,amssymb,amsfonts}
\usepackage{algorithmic}
\usepackage{graphicx}
\usepackage{textcomp}
\usepackage{xcolor}
\usepackage{amsmath,epsfig}
\usepackage{algorithm, algorithmic}
\usepackage{hyperref}
\usepackage{tikz}
\usepackage{subcaption}
\usepackage{amssymb}
\usepackage{graphics}
\usepackage{comment}
\usepackage{multirow}
\makeatletter
\def\ps@IEEEtitlepagestyle{
\def\@oddfoot{\mycopyrightnotice}
\def\@evenfoot{}
}
\def\mycopyrightnotice{
{\footnotesize 978-1-6654-7592-1/22/\$31.00 ~\copyright~2022 IEEE\hfill}
\gdef\mycopyrightnotice{}
}

\def\BibTeX{{\rm B\kern-.05em{\sc i\kern-.025em b}\kern-.08em
    T\kern-.1667em\lower.7ex\hbox{E}\kern-.125emX}}
\begin{document}

\title{Neural Frank-Wolfe Policy Optimization for Region-of-Interest Intra-Frame Coding with HEVC/H.265}

\author{\IEEEauthorblockN{
Yung-Han Ho, Chia-Hao Kao, Wen-Hsiao Peng, Ping-Chun Hsieh } \IEEEauthorblockA{\textit{National Yang Ming Chiao Tung University, Taiwan}}}

\maketitle

\begin{abstract}
This paper presents a reinforcement learning (RL) framework that utilizes Frank-Wolfe policy optimization to solve Coding-Tree-Unit (CTU) bit allocation for Region-of-Interest (ROI) intra-frame coding. Most previous RL-based methods employ the single-critic design, where the rewards for distortion minimization and rate regularization are weighted by an empirically chosen hyper-parameter. Recently, the dual-critic design is proposed to update the actor by alternating the rate and distortion critics. However, its convergence is not guaranteed. To address these issues, we introduce Neural Frank-Wolfe Policy Optimization (NFWPO) in formulating the CTU-level bit allocation as an action-constrained RL problem. In this new framework, we exploit a rate critic to predict a feasible set of actions. With this feasible set, a distortion critic is invoked to update the actor to maximize the ROI-weighted image quality subject to a rate constraint. Experimental results produced with x265 confirm the superiority of the proposed method to the other baselines.
\end{abstract}


\section{Introduction}
\label{sec:introduction} 
Broadly, the task of bit allocation for intra-frame coding is to allocate bits to coding units at certain level in such a way that the reconstructed image quality is maximized subject to a rate constraint. In this paper, we tackle the bit allocation problem at the coding-tree-unit (CTU) level for intra-frame coding with HEVC/H.265. In particular, we consider to weight more heavily the reconstruction quality in regions of interest (ROI). Due to the rate constraint, the ROI prioritization, as well as coding dependencies between CTUs, this problem is in essence a dependent decision-making process. 

Reinforcement learning (RL) lends itself to dependent decision-making. There have been several attempts at applying RL to address bit allocation and rate control for image/video coding. Among the prior works, the single-critic design is most popular. Chen~\textit{et al.}~\cite{chen2018reinforcement} and Zhou~\textit{et al.}~\cite{zhou2020rate} learn RL agents to determine frame-level quantization parameters (QP) for hierarchical B-frame coding and low-delay P-frame coding, respectively. Hu~\textit{et al.}~\cite{hu2018reinforcement} adopt a similar approach for CTU-level bit allocation, in addressing intra-frame rate control. Fu~\textit{et al.}~\cite{360HRL} extend the idea to streaming applications. Li~\textit{et al.}~\cite{li2021task} focus on semantic coding for several computer vision tasks. Ren~\textit{et al.}~\cite{ROIinVVC} tackles ROI-based coding by learning an RL agent for both frame-level and CTU-level bit allocation. These prior works utilize a single reward function usually having a form of $r_D + \lambda r_R$, where $r_D$ and $r_R$ are the distortion and rate rewards, respectively. The design, dubbed the single-critic method, trades off distortion minimization and rate regularization with a fixed hyper-parameter $\lambda$. However, it is difficult to choose a fixed $\lambda$ value that can work well on various videos and bit rates.

Deviating from the single-critic approach, Ho~\textit{et al.}~\cite{ho2021dual} learn two independent critics, one of which estimates $r_D$ and the other $r_R$. They train the RL agent by alternatively using the distortion critic and the rate critic. Specifically, the distortion critic is utilized to update the agent when the rate constraint is satisfied; otherwise, the rate critic is used to train the agent to meet the rate constraint. Though avoiding the use of a fixed $\lambda$, the training convergence is not guaranteed for the dual-critic method.

In this paper, we propose an action-constrained RL framework via Neural Frank-Wolfe Policy Optimization (NFWPO)~\cite{lin2021escaping}, aiming for ROI-based intra-frame coding. Similar to~\cite{ho2021dual}, our scheme is composed of a distortion critic and a rate critic. However, unlike~\cite{ho2021dual}, we apply the rate critic in specifying a state-dependent action feasible set. We then utilize NFWPO together with the distortion critic to identify within the feasible set an action that minimizes the ROI-weighted distortion. The action thus chosen serves as a target for training the agent. We stress that our work also differs from the vanilla NFWPO~\cite{lin2021escaping} in that our action feasible set is dynamically determined via the rate critic rather than predefined. 

To the best of our knowledge, this work presents the first attempt at applying Neural Frank-Wolfe Policy Optimization to address bit allocation and rate control for image/video coding. We demonstrate its effectiveness by taking as an example CTU-level bit allocation for ROI-based intra-frame coding. Experimental results confirm its superiority to the single-critic and dual-critic methods. 


\section{Neural Frank-Wolfe Policy Optimization}
\label{sec:nfwpo}
NFWPO~\cite{lin2021escaping} is an action-constrained RL algorithm. The objective of the action-constrained RL is to maximize the reward-to-go $Q(s,a)$ subject to the feasible actions $\mathcal{C}(s)$:
\begin{equation}
\label{eq:NFWPO_objective}
\arg \max_{a\in \mathcal{C}(s)} Q(s,a),
\end{equation}
where the reward-to-go $Q(s,a)$ is the expected cumulative future reward under the policy $\pi$. In this paper, the policy $\pi(s)$ is implemented by a continuous, deterministic actor network. 

Some prior works \cite{dalal2018safe} deal with the action-constrained RL by including a projection layer at the output of the actor network. The projection layer projects the action onto the feasible set $\mathcal{C}(s)$ by 
\begin{equation}
\label{eq:feasible}
\prod\nolimits_{\mathcal{C}(s)}(a) = \arg \min_{y\in \mathcal{C}(s)} ||y-a||_2,
\end{equation}
where $a=\pi(s)$ is the pre-projection action and $\prod\nolimits_{\mathcal{C}(s)}(a)$ is the post-projection action. Maximizing $Q(s,\prod\nolimits_{\mathcal{C}(s)}(a))$ through gradient ascent may run into the trouble of zero gradients. For example, consider constraining the action to be non-negative with Rectified Linear Unit (ReLU) as the projection layer. The zero-gradient issue occurs during back-propagation when the action falls in the negative region.

To avoid this issue, NFWPO updates the actor network in three consecutive steps. First, it identifies a feasible update direction $\bar{c}(s)$ according to \begin{equation}
\label{eq:c(s)}
\bar c(s)=\arg \max_{c\in \mathcal{C}(s)}\langle c,\nabla_{a} Q(s,a)|_{a=\prod\nolimits_{\mathcal{C}(s)}(\pi(s))}\rangle,
\end{equation} where the operator $\langle a,b\rangle$ takes the inner product of $a$ and $b$. 
Second, it evaluates a reference action $\tilde{a}_s$ by
\begin{equation}
\label{eq:refa}
\tilde{a}_s=\prod\nolimits_{\mathcal{C}(s)}(\pi(s))+\alpha (\bar c(s)-\prod\nolimits_{\mathcal{C}(s)}(\pi(s)),
\end{equation}
where $\alpha$ is the learning rate of NFWPO. Lastly, it learns the actor network $\pi(s)$ through gradient decent by minimizing the squared error between the reference action $\tilde{a}_s$ and $\pi(s)$:
\begin{equation}
\label{eq:upt_actor}
\mathcal{L}_{NFWPO}=(\pi(s)-\tilde a_s)^2.
\end{equation}
The zero-gradient issue is circumvented during training as the projection layer is not involved in the back-propagation.

\section{Proposed Method}
\label{sec:pmethod}
The objective of the CTU-level bit allocation for ROI-based intra-frame coding is to minimize the ROI-weighted distortion subject to a rate constraint. This is achieved by choosing properly a quantization parameter (QP) for every CTU in an intra-frame. In symbols, we have
\begin{equation}
\label{eq:objective}
\arg \min_{\{QP_i\}}\sum_{i=1}^N D_i(QP_i)\text{ s.t.}\sum_{i=1}^N R_i(QP_i) \leq R_f,
\end{equation}
where $QP_i$ indicates the QP for the i-th CTU, $N$ denotes the number of CTUs in a frame, $D_i(QP_i)$ is the distortion of CTU $i$ encoded with $QP_i$, $R_i(QP_i)$ is the number of encoded bits of CTU $i$, and $R_f$ is the frame-level bit budget. To achieve ROI-based intra-frame coding, the distortions of CTUs in ROI regions are weighted more heavily.

We formulate our task as an action-constrained RL problem. When determining $QP_i$ to encode CTU $i$, we view the minimization of the cumulative distortion in Eq.~\eqref{eq:objective} as the maximization of the reward-to-go in Eq.~\eqref{eq:NFWPO_objective}. To meet the rate constraint in Eq.~\eqref{eq:objective}, we limit $QP_i$ to a state-dependent feasible set $\mathcal{C}(s_i)$, which is specified by a rate critic (Section~\ref{subsec:RC-NFWPO}). Eq.~\eqref{eq:objective} is then transformed into 
\begin{equation}
\label{eq:reform_objective}
\arg \max_{QP_i\in \mathcal{C}(s_i)} Q (s_i,QP_i), \forall i,
\end{equation} which takes the same form as Eq.~\eqref{eq:NFWPO_objective} and allows us to train the actor with NFWPO (Section~\ref{sec:nfwpo}).   

\subsection{System Overview}
\label{subsec:oarch}
\begin{figure}[t!]
\vspace*{-2mm}
\centering
\includegraphics[width=\linewidth]{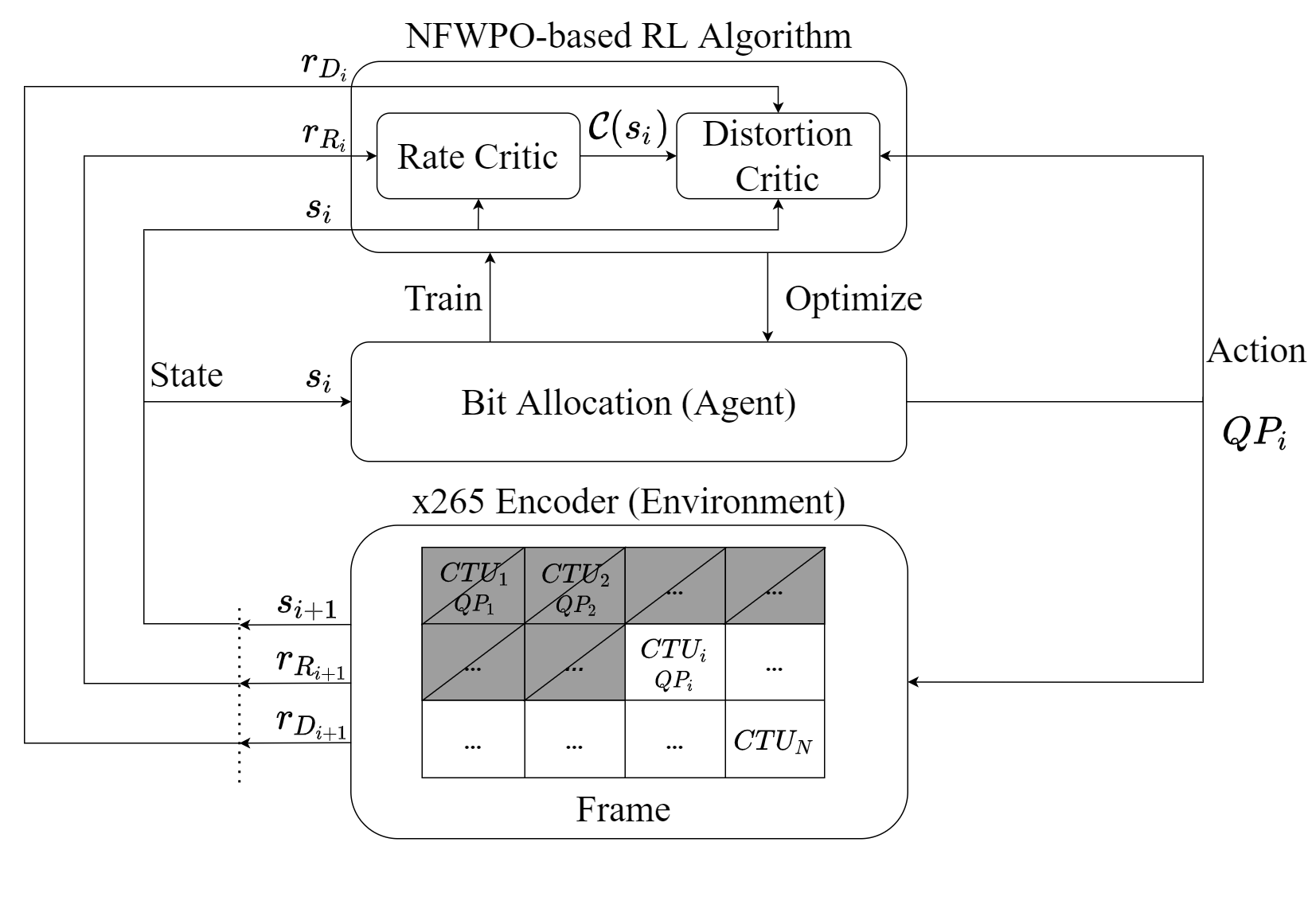}
\vspace*{-7mm}
\caption{The proposed NFWPO-based RL framework for CTU-level bit allocation.}
\label{fig:RL}
\vspace*{-2mm}
\end{figure}

Fig.~\ref{fig:RL} illustrates our action-constrained RL framework. When encoding CTU $i$, a state $s_i$ is first evaluated. Taking this state as input, our RL agent outputs an action $QP_i$ (Section \ref{subsec:state}). The x265 codec then encodes CTU $i$ with $QP_i$. After encoding CTU $i$, we evaluate a distortion reward $r_{D_{i}}$ and a rate reward $r_{R_{i}}$ (Section \ref{subsec:rewards}). These steps are repeated until all the CTUs in a frame are encoded.

At training time, the agent interacts with x265 by encoding every frame as an episodic task. We utilize the distortion and rate critics to predict the distortion reward-to-go $Q_D(s, QP)$ and the rate reward-to-go $Q_R(s, QP)$, respectively. The rate critic, which predicts the rate deviation from $R_f$ at the end of encoding a frame, enables us to specify a feasible set $\mathcal{C}(s_i)$ of $QP_i$. The distortion critic, which estimates the cumulative distortion, guides the agent to minimize the total distortion. That is, $Q_D$ will play the role of $Q$ in Eq.~\eqref{eq:reform_objective}.

\begin{table}[tb]
\setlength{\abovecaptionskip}{0cm}
\begin{small}
\begin{center}
\caption{State Definition}
\label{tab:state}
\scalebox{1.0}{
\begin{tabular}{|c|c|}
\hline
\multicolumn{1}{c}{} & \multicolumn{1}{c}{Components}\\
\hline
\multicolumn{1}{l}{\textbf{1}} & \multicolumn{1}{l}{Variance of the current CTU}  \\
\multicolumn{1}{l}{\textbf{2}} & \multicolumn{1}{l}{Gradient of the current CTU}  
\\
\multicolumn{1}{l}{\textbf{3}} & \multicolumn{1}{l}{Average of variances over remaining CTUs in the frame}  \\
\multicolumn{1}{l}{\textbf{4}} & \multicolumn{1}{l}{Average of gradients over remaining CTUs in the frame}  \\
\multicolumn{1}{l}{\textbf{5}} & \multicolumn{1}{l}{Percentage of the outstanding bits} \\
\multicolumn{1}{l}{\textbf{6}} & \multicolumn{1}{l}{Percentage of the remaining CTUs in the frame} \\
\multicolumn{1}{l}{\textbf{7}} & \multicolumn{1}{l}{Base QP  (see Section~\ref{subsec:state})} \\
\multicolumn{1}{l}{\textbf{8}} & \multicolumn{1}{l}{Bit budget of the current frame} \\
\multicolumn{1}{l}{\textbf{9}} & \multicolumn{1}{l}{Current CTU ROI indicator} \\
\multicolumn{1}{l}{\textbf{10}} & \multicolumn{1}{l}{Percentage of remaining ROI CTUs in the frame} \\
\hline
\end{tabular}
}
\end{center}
\end{small}
\vspace*{-5mm}
\end{table}

\subsection{States and Actions}
\label{subsec:state}
Inspired by~\cite{hu2018reinforcement}, we provide our (hand-crafted) state representation in Table~\ref{tab:state}, which serves as the basis for the agent to output the action. 

The action of our RL agent indicates the QP difference (also known as the delta QP) from the base QP. That is, the final QP value is the sum of the delta and the base QPs (i.e. $QP_i = \text{delta QP} + \text{base QP}$). The base QP is designed to reduce the search space of our agent; its value depends on the rate point. 

\subsection{Rewards}
\label{subsec:rewards}
We specify two immediate rewards: the distortion reward $r_{D_i}$ and the rate reward $r_{R_i}$. We define $r_{D_i}$ as
\begin{equation}
\label{eq:rewardd}
r_{D_i}= 
\begin{cases}
    -\text{D}_i\cdot w, & \text{if } \text{CTU}_i \text{ is ROI}; \\
    -\text{D}_i, & \text{otherwise}, \\
\end{cases}
\end{equation}
where $\text{D}_i$ is the mean-squared error (MSE) of CTU $i$ when encoded with the QP value chosen by the agent, and $w\geq1$ is used to weight more heavily the distortions of ROI CTUs.

To construct the feasible set that addresses the rate constraint, the immediate rate reward $r_{R_i}$ is designed as
\begin{equation}
\label{eq:Rreward}
r_{R_i}= 
\begin{cases}
    \frac{-|R_f-\sum_{t=1}^N R_t(QP_t)|}{R_f}, & \text{if } i=N; \\
    0, & \text{otherwise}. \\
\end{cases}
\end{equation}
The $\sum_{i=1}^N r_{R_i}$ represents the negative absolute deviation of the coding bit rate from the target $R_f$ in percentage terms. 

With $r_{D_i}$ and $r_{R_i}$, the distortion and rate reward-to-go's $Q_D(s,QP)$ and $Q_R(s,QP)$ in Section~\ref{subsec:oarch} are given by
\begin{equation}
\label{eq:D_reward-to-go}
Q_D(s_i,QP_i) = E_{(s_t,QP_t) \sim \pi}[\sum\nolimits_{t=i}^N \gamma^{t-i} r_{D_t}]
\end{equation}
\begin{equation}
\label{eq:R_reward-to-go}
Q_R(s_i,QP_i) = E_{(s_t,QP_t) \sim \pi}[\sum\nolimits_{t=i}^N \gamma^{t-i} r_{R_t}],
\end{equation}
where $\gamma$ is the discount factor. Particularly, these reward-to-go functions are approximated by the distortion and rate critics.

\begin{figure}[t]
\centering
\vspace{-2mm}
\includegraphics[width=\linewidth]{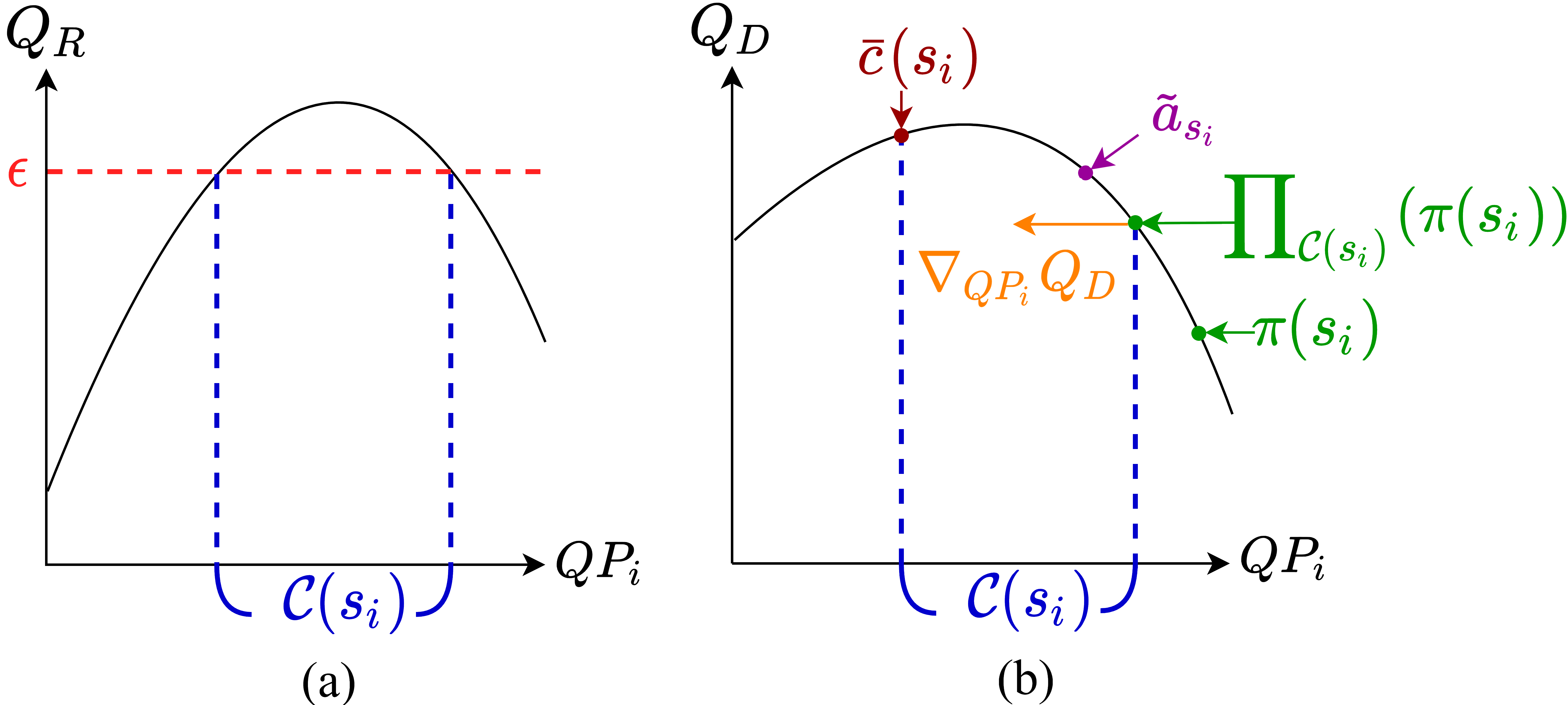}
\caption{Illustration of (a) the feasible set $\mathcal{C}(s)$, and (b) the reference action $\tilde{a}_s$.}
\vspace{-5mm}
\label{fig:adapted_NFWPO}
\end{figure}

\subsection{NFWPO-based RL for CTU-level Bit Allocation}
\label{subsec:RC-NFWPO}
This section presents how we use the two critic networks to implement NFWPO. First, we identify the feasible set $\mathcal{C}(s_i)$ for CTU $i$ by the rate critic. To satisfy $R_f$, $\mathcal{C}(s_i)$ includes the QP values $QP_i$ that the rate reward-to-go $Q_R(s_i,QP_i)$ is greater than or equal to a threshold $\epsilon$ (see Fig.~\ref{fig:adapted_NFWPO} (a)):
\begin{equation}
\label{eq:rcfsetqr}
\mathcal{C}(s_i)=\{QP_i|Q_R(s_i,QP_i)\geq\epsilon\}.
\end{equation}
According to Eqs.~\eqref{eq:Rreward} and \eqref{eq:R_reward-to-go}, $\mathcal{C}(s_i)$ contains QP values that ensure the absolute rate deviation is capped by $\epsilon$. Specifically, we discretize these QP values by querying the rate critic $Q_R$ with a discrete step of 0.1 in the range of delta QP (i.e. $QP_{i} = \{\text{base QP}-10, \text{base QP}-9.9, ..., \text{base QP}+10$\}).

Given the feasible set $\mathcal{C}(s_i)$, we follow the procedure in Section \ref{sec:nfwpo} to generate the reference action $\tilde{a}_s$ (Fig.~\ref{fig:adapted_NFWPO} (b)). If the actor output $\pi(s_i)$ is outside of the feasible set, it will be projected onto the feasible set to reach $\Pi_{\mathcal{C}(s_i)}(\pi(s_i))$. We then derive $\bar{c}(s_i)$ based on Eq.~\eqref{eq:c(s)}, where we replace $Q(s,a)$ with $Q_D(s,a)$, and get $\tilde{a}_s$ by Eq.~\eqref{eq:refa}. Finally, we update the actor network via Eq.~\eqref{eq:upt_actor}. Algorithm \ref{alg:A_NFWPO} summarizes the proposed method.

\setlength{\textfloatsep}{1pt}
\begin{algorithm}[t]
\begin{footnotesize}
\caption{The proposed NFWPO-based RL algorithm}
\label{alg:A_NFWPO}
\begin{algorithmic}[1]
\STATE {Randomly initialize critics $Q_D(s,a|w_D), Q_R(s,a|w_R)$, and actor $\pi(s|\theta)$} with weights $w_D,w_R,$ and $\theta$
\STATE {Initialize target networks $Q_D'(s,a|w_D'), Q_R'(s,a|w_R')$, and $\pi'(s|\theta')$} with weights $w_D' \xleftarrow{} w_D,w_R' \xleftarrow{}w_R,$ and $\theta' \xleftarrow{}\theta$
\STATE {Initialize replay buffer $R$}
\STATE {\textbf{for} episode = $1$ to $M$ \textbf{do}}
\STATE {\quad Initialize a random noise process $\mathcal{N}$ for action exploration}
\STATE {\quad Evaluate initial state $s_1$}
\STATE {\quad \textbf{for} CTU $i = 1$ to $N$ in a frame \textbf{do}}
\STATE {\quad\quad Set $a_i = \pi(s_i|\theta)+\mathcal{N}_i$ }
\STATE {\quad\quad Encode CTU $i$ with $QP=a_i$}
\STATE {\quad\quad Evaluate the immediate rewards $r_{D_i}$, $r_{R_i}$ and the new state $s_{i+1}$}
\STATE {\quad\quad Store transition $(s_i, a_i, r_{D_i}, r_{R_i}, s_{i+1})$ in $R$}
\STATE {\quad \textbf{end for}}
\STATE {\quad Sample $\mathcal{B}$ transitions $(s_{b}, a_{b}, r_{D_{b}}, r_{R_{b}}, s_{b+1})$ from $R$}
\STATE {\quad Set $y_{D_b} = r_{D_b}+\gamma Q_D'(s_{b+1},\pi'(s_{b+1}|\theta')|w_D')$ }
\STATE {\quad Update $Q_D$ by minimizing $\mathcal{L}=\frac{1}{\mathcal{B}}\Sigma_b{(y_{D_b}-Q_D(s_b,a_b|w_D))^2}$}
\STATE {\quad Set $y_{R_b} = r_{R_b}+\gamma Q_R'(s_{b+1},\pi'(s_{b+1}|\theta')|w_R')$ }
\STATE {\quad Update $Q_R$ by minimizing $\mathcal{L}=\frac{1}{\mathcal{B}}\Sigma_b{(y_{R_b}-Q_R(s_b,a_b|w_R))^2}$}
\STATE {\quad \textbf{for} each state s $\in \mathcal{B}$ \textbf{do}}
\STATE {\quad\quad Identify the feasible set $\mathcal{C}(s)$ by Eq.~\eqref{eq:rcfsetqr}}
\STATE {\quad\quad Obtain $\bar{c}(s)$ by replacing $Q$ with $Q_D$ in Eq.~\eqref{eq:c(s)}}
\STATE {\quad\quad Obtain the reference action $\tilde{a}_s$ by Eq.~\eqref{eq:refa}}
\STATE {\quad\quad Update the actor network $\pi$ by Eq.~\eqref{eq:upt_actor}}
\STATE {\quad \textbf{end for}}
\STATE {\quad Update target networks $Q_D', Q_R', $and $\pi'$}
\STATE {\textbf{end for}}
\end{algorithmic}
\end{footnotesize}
\end{algorithm}
\section{Experimental Results}
\label{sec:eresults}

\begin{table*}[t]
\centering
\caption{Comparison of Rate Deviations (at the lowest rate point) and BD-rates in different ROI settings (x265 as anchor)}
\vspace{-2mm}
\label{tab:exp_table}
{\renewcommand{\arraystretch}{1.2}
\begin{tabular}{|c|c|c||ccc||ccc||l|c|}
\hline
\multirow{2}{*}{\textbf{\#}} & \multirow{2}{*}{\textbf{Test Set}}       & \multirow{2}{*}{\textbf{\begin{tabular}[c]{@{}c@{}}ROI\\ Condition\end{tabular}}} & \multicolumn{3}{c||}{\textbf{Rate Deviation (\%)}}                                  & \multicolumn{3}{c||}{\textbf{BD-Rate (\%)}}                                & \multicolumn{1}{c|}{\multirow{2}{*}{\textbf{ROI Details}}} & \multirow{2}{*}{\textbf{\begin{tabular}[c]{@{}c@{}}Avg. ROI\\size (CTU)\end{tabular}}} \\ \cline{4-9}
                             &                                          &                                                                                   & \multicolumn{1}{c|}{single} & \multicolumn{1}{c|}{Dual} & Ours          & \multicolumn{1}{c|}{single}          & \multicolumn{1}{c|}{Dual}   & Ours           & \multicolumn{1}{c|}{}                                      &                                                                                         \\ \hline
1                            & \textit{\textbf{DAVIS}}                  & Regular                                                                           & \multicolumn{1}{c|}{6.39}   & \multicolumn{1}{c|}{1.57} & \textbf{0.96} & \multicolumn{1}{c|}{\textbf{-20.09}} & \multicolumn{1}{c|}{-15.98} & -18.79         & CTUs corresponding to all objects in mask                  & 10.7                                                                                    \\ \hline
2                            & \textit{\textbf{COCO1}}                  & Regular                                                                           & \multicolumn{1}{c|}{19.95}  & \multicolumn{1}{c|}{4.92} & \textbf{3.26} & \multicolumn{1}{c|}{-6.31}           & \multicolumn{1}{c|}{-6.53}  & \textbf{-6.63} & CTUs corresponding to randomly picked objects              & 15.9                                                                                    \\ \hline
3                            & \multirow{2}{*}{\textit{\textbf{COCO2}}} & Small ROI                                                                         & \multicolumn{1}{c|}{4.59}   & \multicolumn{1}{c|}{2.45} & \textbf{1.50} & \multicolumn{1}{c|}{-2.92}           & \multicolumn{1}{c|}{-5.51}  & \textbf{-7.34} & CTUs corresponding to the smallest object                  & 2.5                                                                                     \\ \cline{1-1} \cline{3-11} 
4                            &                                          & Large ROI                                                                         & \multicolumn{1}{c|}{48.59}  & \multicolumn{1}{c|}{6.92} & \textbf{5.91} & \multicolumn{1}{c|}{6.42}            & \multicolumn{1}{c|}{6.47}   & \textbf{5.21}  & Inversion of Small ROI                                     & 37.5                                                                                    \\ \hline
\end{tabular}
}
\vspace{-5mm}
\end{table*}

\subsection{Settings and Training Details}
\label{subsec:esettings}

We conduct our experiments using x265 under the all-intra configuration (\textit{-{}-keyint 1}). All the sequences in our experiments are resized to $512 \times 320$ with CTU size $64 \times 64$, resulting in 40 CTUs per frame. The experiments follow the same frame-level bit allocation as x265. That is, we encode every sequence with fixed QPs 22, 27, 32, and 37 to establish the frame-level bit budget $R_f$, and turning on \textit{-{}-tune psnr} to optimize for PSNR. 


Two datasets are used in our experiments, DAVIS 2017 TrainVal~\cite{Pont-Tuset_arXiv_2017} and COCO 2017 validation~\cite{lin2014microsoft}, both of which provide ground truth object masks. For training, we use 64 sequences from DAVIS, and the number and position of ROI CTUs are sampled randomly. 

At test time, we experiment with 4 settings on 3 video test sets. We utilize the ground truth masks to specify CTU-level ROI by defining ROI as CTUs that overlap with the selected object mask. In Table~\ref{tab:exp_table}, the \textit{regular ROI} setting is tested on \textbf{\textit{DAVIS}} test set (\#1), which consists of 20 sequences from DAVIS dataset with ROI specified by the ground truth object masks, and on \textbf{\textit{COCO1}} (\#2), formed by 1600 images from COCO dataset with ROI given by the object masks of randomly chosen categories. We experiment with the \textit{small ROI} and the \textit{large ROI} settings on \textbf{\textit{COCO2}} test set (\#3, 4), respectively. The \textbf{\textit{COCO2}} test set is composed of 950 images collected from COCO dataset. The \textit{small ROI} setting is tested on the selected images, where ROI corresponds to the smallest object that covers no more than 5 CTUs. The \textit{large ROI} setting simply inverts the ROI specification of the \textit{small ROI} setting. 

To train the actor and critic networks, we choose $\alpha=0.05$ in Eq.~\eqref{eq:refa}, the ROI weighting parameter $w=10$ in Eq.~\eqref{eq:rewardd}, the learning rate to be $0.001$, and the 3-step temporal difference method.  The base QPs are set to $QP_l-3$, where $QP_l$ are 22, 27, 32, and 37. The delta QP ranges from -10 to 10. The threshold $\epsilon$ in Eq.~\eqref{eq:rcfsetqr} is set to $-0.05$, allowing for a maximum rate deviation of $\pm5\%$. For a fair comparison, the same rate tolerance is applied to the single- and dual-critic methods. 

\subsection{Rate-distortion Performance and Rate Deviations}
\label{subsec:rd}
Table~\ref{tab:exp_table} presents BD-rates (in terms of PSNR-YUV where the Y,U,V distortions are weighted in proportional to 6:1:1) and rate deviations, with x265 serving as anchor. In particular, the ROI-weighted MSE is evaluated according to 
\begin{equation}
\label{eq:weighted_MSE}
\text{ROI-Weighted MSE} =\frac{\text{MSE}_\text{ROI}\times 10+\text{MSE}_\text{NROI}}{N_\text{ROI}\times 10+N_\text{NROI}},
\end{equation}
where $\text{MSE}_\text{ROI}$ and $N_\text{ROI}$ are the sum of MSEs and the number of ROI CTUs, respectively; $\text{MSE}_\text{NROI}$ and $N_\text{NROI}$ are those of non-ROI CTUs. When reporting the average absolute rate deviations from the frame-level bit budget $R_f$, any deviation within $\pm5\%$ of the $R_f$ is regarded as $0\%$ to reflect our $\pm5\%$ tolerance. 

From Table~\ref{tab:exp_table}, we see that our scheme achieves the smallest rate deviation under all the ROI settings, compared with the single-critic~\cite{hu2018reinforcement} and dual-critic~\cite{ho2021dual} methods. In contrast, the rate deviation of the single-critic method is seen to be as high as 50\% under the \textit{Large ROI} setting. This is attributed to the use of a fixed $\lambda$ for combining the distortion and the rate rewards ($r_D+\lambda r_R$). In the present case, the cumulative distortion reward may change drastically with the number of ROI CTUs (cp. Eq.~\eqref{eq:rewardd}). This makes it difficult to identify a fixed $\lambda$ that can work well under various ROI settings. We observe that the policy learned by the single-critic method~\cite{hu2018reinforcement} achieves the best test result only when the number of ROI CTUs is close to the training average (i.e. 20) and the test sequences share similar characteristics to the training data (i.e. DAVIS). When tested with \textit{large ROI}, the single-critic method chooses too low a QP for ROI CTUs, leading to large rate overshooting (see \#2, 4); in the other extreme with \textit{small ROI}, it assigns too high a QP to non-ROI CTUs, resulting in poor rate-distortion performance (\#3).
In terms of BD-rate performance, our method outperforms the single-critic~\cite{hu2018reinforcement} and the dual-critic~\cite{ho2021dual} methods in most of the settings. Even though the single-critic method shows slightly better BD-rate results in setting \#1 (where the ROI setting and sequences are more similar to those used for training), it exhibits poor performance in the other settings (e.g. \#3, 4), which underlines its poor generalization performance. In comparison, our method shows more consistent results across different settings.

\begin{figure}[t]
\vspace*{-1mm}
\centering
\includegraphics[width=0.94\linewidth]{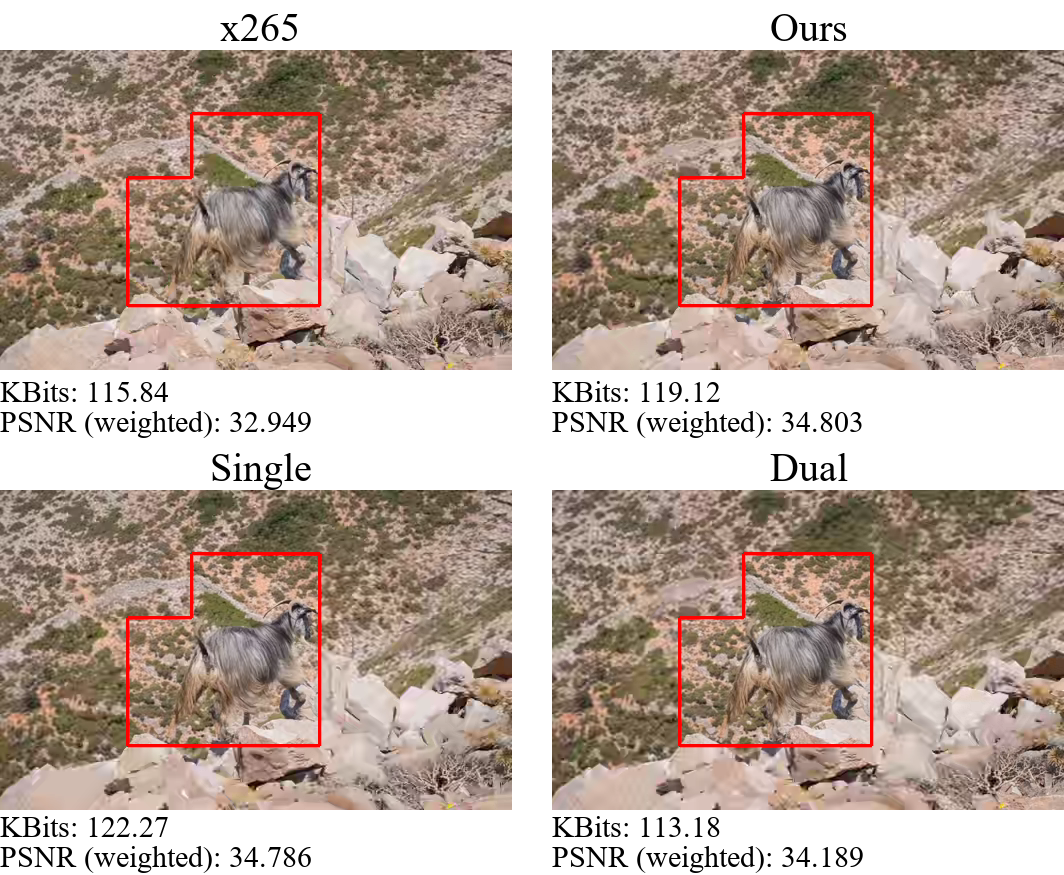}
\caption{Subjective quality comparison with ROI highlighted.}
\label{fig:subjectiv}
\end{figure}

\begin{figure}[t]
\vspace*{-4mm}
\centering
\begin{center}
\includegraphics[width=0.94\linewidth]{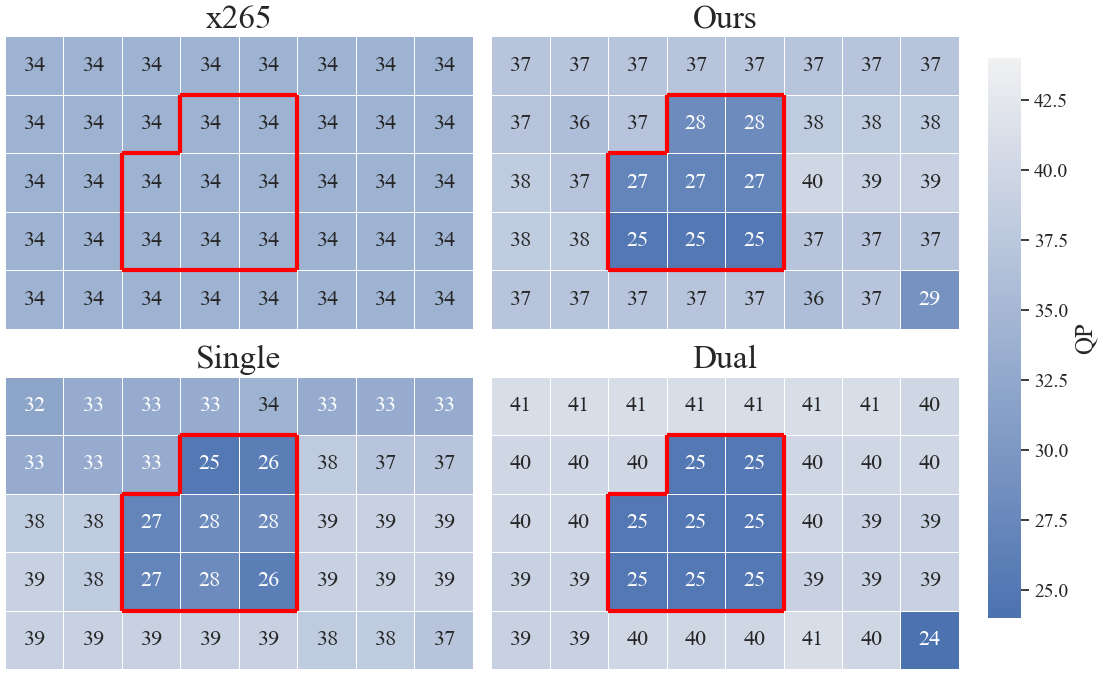}
\end{center}
\vspace*{-1mm}
\caption{Visualization of QP assignment.}
\label{fig:heatmap}
\end{figure}

Fig.~\ref{fig:subjectiv} further presents a subjective quality comparison. As compared to the other methods, ours preserves more texture details in ROI and shows less blocking artifacts. Fig.~\ref{fig:heatmap} visualizes the corresponding QP assignment. Our method assigns lower QPs in ROI CTUs, which is in stark contrast to x265. One thing to note is that both the dual-critic method and ours choose a low QP for the last non-ROI CTU. This is resulted from the higher QPs assigned to previous CTUs. To meet the rate constraint, a lower QP is chosen for the last CTU.

\section{Conclusion}
\label{sec:conclusion}
This paper introduces a NFWPO-based RL framework for ROI-based intra-frame coding with HEVC/H.265. It overcomes the empirical choice of the hyper-parameter in the single-critic method and the convergence issue of the dual-critic method. It outperforms these two baselines, demonstrating better ability to generalize to various ROI settings.

\bibliographystyle{IEEEtran}
\bibliography{ref}

\end{document}